\documentclass[twocolumn]{aastex631}

\usepackage{graphicx}
\usepackage[mathscr]{euscript}
\usepackage{float}
\usepackage{natbib}
\usepackage{CJK}

\begin{document}
\begin{CJK*}{UTF8}{gbsn}

\title{Resolving the Young 2~Cygni Run-away Star into a Binary using iLocater} 
\shorttitle{2 Cygni AO Imaging}

\author[0000-0003-0800-0593]{Justin R. Crepp}
\affiliation{University of Notre Dame, Department of Physics and Astronomy, Notre Dame, IN, USA}
%\altaffiliation{aka Clark Kent}  % Make a special note about an author. 

\author[0000-0002-1503-2852]{Jonathan Crass}
\affiliation{The Ohio State University, Department of Astronomy, Columbus, OH, USA} 
\affiliation{University of Notre Dame, Department of Physics and Astronomy, Notre Dame, IN, USA}

\author[0000-0002-3047-9599]{Andrew J. Bechter}
%\affiliation{Northrop Grumman Corporation, 1 Space Park Drive, Redondo Beach, CA, USA}
\affiliation{University of Notre Dame, Department of Physics and Astronomy, Notre Dame, IN, USA}

\author[0009-0000-4650-2266]{Brian L. Sands}
\affiliation{University of Notre Dame, Department of Physics and Astronomy, Notre Dame, IN, USA}

\author{Ryan Ketterer}
\affiliation{University of Notre Dame, Department of Physics and Astronomy, Notre Dame, IN, USA}

\author{David King}
\affiliation{Durham University, Department of Physics, Durham, UK}
\affiliation{University of Cambridge, Institute of Astronomy, Cambridge, UK}

\author[0000-0003-0121-5815]{Derek Kopon}
%\affiliation{MIT Lincoln Laboratory, Lexington, MA, USA}
%\affiliation{Harvard Smithsonian Center for Astrophysics, Cambridge, MA, USA}

\author{Randall Hamper}
\affiliation{University of Notre Dame, Department of Physics and Astronomy, Notre Dame, IN, USA}

\author{Matthew Engstrom}
\affiliation{University of Notre Dame, Department of Physics and Astronomy, Notre Dame, IN, USA}

\author{James E. Smous}
\affiliation{University of Notre Dame, Department of Physics and Astronomy, Notre Dame, IN, USA}

\author[0000-0001-8725-8730]{Eric B. Bechter}
\affiliation{University of Notre Dame, Department of Physics and Astronomy, Notre Dame, IN, USA}

\author[0000-0002-8518-4640]{Robert Harris}
\affiliation{Durham University, Department of Physics, Durham, UK}
%\affiliation{Centre for Advanced Instrumentation, Department of Physics, Durham University, South Road, Durham, UK}

\author[0000-0002-5099-8185]{Marshall C. Johnson}
\affiliation{The Ohio State University, Department of Astronomy, Columbus, OH, USA}

\author{Nicholas Baggett}
\affiliation{University of Notre Dame, Department of Physics and Astronomy, Notre Dame, IN, USA}

\author{Shannon Dulz}
\affiliation{University of Notre Dame, Department of Physics and Astronomy, Notre Dame, IN, USA}

\author{Michael Vansickle}
\affiliation{University of Notre Dame, Department of Physics and Astronomy, Notre Dame, IN, USA}

% ALPHABETIZE AFTER HERE

\author{Al Conrad}
\affiliation{Large Binocular Telescope Observatory, Tucson, AZ, USA}

\author[0000-0002-2314-7289]{Steve Ertel}
\affiliation{University of Arizona, Department of Astronomy, Tucson, AZ, USA}
\affiliation{Large Binocular Telescope Observatory, Tucson, AZ, USA}

\author[0000-0003-0395-9869]{B. Scott Gaudi}
\affiliation{The Ohio State University, Department of Astronomy, Columbus, OH, USA}

\author[0000-0002-1954-4564]{Philip Hinz} 
\affiliation{UC Santa Cruz, Department of Astronomy and Astrophysics, Santa Cruz, CA, USA}
\affiliation{Large Binocular Telescope Observatory, Tucson, AZ, USA}

\author[0000-0002-2387-5489]{Marc Kuchner}
\affiliation{Goddard Space Flight Center, Greenbelt, MD, USA}

\author{Manny Montoya}
\affiliation{Large Binocular Telescope Observatory, Tucson, AZ, USA}

\author{Eleanya Onuma} 
\affiliation{Goddard Space Flight Center, Greenbelt, MD, USA}

\author{Melanie Ott}
\affiliation{Goddard Space Flight Center, Greenbelt, MD, USA}

\author[0000-0003-1435-3053]{Richard Pogge}
\affiliation{The Ohio State University, Department of Astronomy, Columbus, OH, USA}

\author{Gustavo Rahmer}
%\affiliation{GMT}
\affiliation{Large Binocular Telescope Observatory, Tucson, AZ, USA}

\author{Robert Reynolds}
\affiliation{Large Binocular Telescope Observatory, Tucson, AZ, USA}

\author[0000-0002-4046-987X]{Christian Schwab}
\affiliation{Macquarie University, School of Mathematical and Physical Sciences, Sydney, AU}

\author[0000-0002-2805-7338]{Karl Stapelfeldt}
\affiliation{Goddard Space Flight Center, Greenbelt, MD, USA}
\affiliation{Jet Propulsion Laboratory, California Institute of Technology, Pasadena, CA, USA}

\author{Joseph Thomes}
\affiliation{Goddard Space Flight Center, Greenbelt, MD, USA}

\author{Amali Vaz}
\affiliation{Large Binocular Telescope Observatory, Tucson, AZ, USA}

\author[0000-0002-4361-8885]{Ji Wang(王吉)}
\affiliation{The Ohio State University, Department of Astronomy, Columbus, OH, USA}

\author[0000-0001-6567-627X]{Charles E. Woodward}
\affiliation{University of Minnesota, Minnesota Institute for Astrophysics, Minneapolis, MN, USA}

\correspondingauthor{Justin Crepp}
\email{jcrepp@nd.edu} 

\begin{abstract}
Precision radial velocity (RV) spectrographs that use adaptive optics (AO) show promise to advance telescope observing capabilities beyond those of seeing-limited designs. We are building a spectrograph for the Large Binocular Telescope (LBT) named iLocater that uses AO to inject starlight directly into single mode fibers (SMF). iLocater's first acquisition camera system (the `SX' camera), which receives light from one of the 8.4m diameter primary mirrors of the LBT, was initially installed in summer 2019 and has since been used for several commissioning runs. We present results from first-light observations that include on-sky measurements as part of commissioning activities. Imaging measurements of the bright B3IV star 2 Cygni ($V=4.98$) resulted in the direct detection of a candidate companion star at an angular separation of only $\theta = 70$ mas. Follow-up AO measurements using Keck/NIRC2 recover the candidate companion in multiple filters. An $R\approx1500$ miniature spectrograph recently installed at the LBT named ``Lili'' provides spatially resolved spectra of each binary component, indicating similar spectral types and strengthening the case for companionship. Studying the multiplicity of young runaway star systems like 2 Cygni ($36.6 \pm 0.5$ Myr) can help to understand formation mechanisms for stars that exhibit anomalous velocities through the galaxy. This on-sky demonstration illustrates the spatial resolution of the iLocater SX acquisition camera working in tandem with the LBT AO system; it further derisks a number of technical hurdles involved in combining AO with Doppler spectroscopy.\vspace{0.3in}
\end{abstract} 

% overcome many of the limitations of seeing-limited designs

\section{Introduction}\label{sec:intro}

Radial velocity (RV) spectrographs have traditionally been installed on seeing-limited telescopes, or telescope ports that do not correct for atmospheric turbulence \citep{fischer_16}. This design decision restricts spatial resolution, which in turn places practical limitations on spectral resolution. The ability to measure precise Doppler shifts is predicated on the quality, compactness, and signal-to-noise ratio of delivered images \citep{ebechter_21}. On a large telescope, adaptive optics (AO) improves spatial resolution by an order of magnitude relative to typical seeing conditions (100 mas versus 1000 mas). Thus, diffraction-limited imaging not only enhances both spatial and spectral resolution, but also enables the use of smaller core fibers --- including those that propagate only a single spatial mode. This latter effect reduces the size of the instrument, allowing for the use of intrinsically stable materials that improve thermal stability, while eliminating fiber modal noise entirely \citep{robertson_12,schwab_12,crepp_14}.\footnote{See \cite{abechter_20_pol} for a discussion of polarization noise.}

Of particular concern are Doppler uncertainties introduced by stellar absorption line asymmetries \citep{cegla_13,cegla_18,cegla_19}. Subtle changes in the line profiles of stellar spectra are caused by surface inhomogeneities (spots, plage, faculae, granulation, etc.). These higher-order, time-dependent and wavelength-dependent variations can only be measured at spectral resolutions above $R\approx150,000$ at high signal-to-noise ratio \citep{strassmeier_15,davis_17}. The most stable and precise RV spectrographs are limited by such effects \citep{dumusque_14,plavchan_15,dumusque_18,crass_21_report}.

Spatial resolution also impacts the type of science cases that can be pursued. For example, unwanted spectral contamination from an unresolved binary or unrelated background star can cause large systematic errors that vary in time, even when using optical fibers \citep{wright_13}. As a result, ground-based exoplanet studies have primarily concentrated on observations of single stars to date. Overcoming this observational bias to study multi-star systems would help challenge and substantiate theoretical models of planet formation and evolution \citep{moe_20}. 

Follow-up observations of transiting planets discovered from space missions like \emph{Kepler} and \emph{TESS} would also benefit from diffraction-limited RV spectroscopy \citep{borucki_13,ricker_15}. Unresolved binaries modify the radius and density estimates of transiting planets \citep{howell_12,ciardi_13,schwamb_13,bechter_14,eastman_16}. Further, in the case of resolved binaries or chance-aligned background stars, uncertainty can remain regarding which star hosts which planet(s) \citep{morton_16}.

iLocater is a diffraction-limited Doppler spectrograph being developed for the LBT \citep{crepp_16}. The instrument receives an AO-corrected beam and injects starlight directly into single mode fibers (SMFs), one for each 8.4m diameter primary mirror plus a calibration fiber \citep{abechter_16,abechter_20_smf,crass_21}. By achieving high spatial and spectral resolution simultaneously, iLocater shows promise to study planets in close-separation binaries. It will also advance our understanding of the masses, densities, orbits, and spin-orbit alignment of transiting planets, particularly around late-type stars \citep{ebechter_19_drp,ebechter_19_h4rg}. For general astrophysics, the instrument may further be used to study the solar neighborhood involving: crowded fields (young stellar cluster dynamics); weak absorption lines (composition of low-metallicity stars); line blanketing (substellar object atmospheric observations); and other applications \citep{crepp_16}. 

iLocater will use two separate acquisition camera systems, SX (left) and DX (right), to acquire light collected from each LBT primary mirror. Following delivery of hardware and confirmation of laboratory performance, the first iLocater acquisition camera system (SX) and fiber injection system was installed on the telescope in June 2019. Day-time engineering activities allowed for co-alignment to the LBT Interferometer (LBTI) system, which provides an AO-corrected beam to the instrument from the Single-conjugate AO Upgrade for the LBT (SOUL) \citep{hinz_16,pinna_16}. Further details of the SX acquisition camera design and technical engineering results can be found in \cite{crass_21}. Once the remaining instrument modules are commissioned, forthcoming observations will use both the SX and DX acquisition cameras \citep{crass_22_design}. The light will be coupled into two different single mode fibers (SMF), one located at each AO port (and a third fiber for calibration), allowing for simultaneous collection of spectra from closely separated sources using the same spectrograph \citep{abechter_20_smf,crass_22_design}.

The star 2~Cygni was observed in July 2019 as part of first-light experiments from the commissioning of iLocater's SX acquisition camera. 2~Cygni has an elevated Gaia renormalized unit weight error (RUWE) value of 1.468, suggesting that it may be a multi-star system \citep{makarov_05,gaia_dr3_21}. Serendipitously observed as a nearby bright star to diagnose hardware, imaging observations in the near-infrared ($\lambda=0.927-0.960 \; \mu$m) using the LBT AO system spatially resolved 2~Cygni into two distinct components. The very small measured angular separation of 70 mas ($\theta \approx 3 \; \lambda/D$) suggests that the candidate may be a companion star. Follow-up imaging using Keck AO in July 2022 and spectroscopy using the ``Lili" instrument at LBT in May 2024 corroborate the hypothesis that 2 Cygni may be a stellar binary system. These LBT commissioning observations demonstrate a key capability of the iLocater instrument, which is designed to operate at the diffraction limit.

% were conducted over three half-nights in summer 2019
% including two nights of consecutive observations of 2~Cygni (09-July UT, 10-July UT)

In this paper, we describe: the literature surrounding 2~Cygni ($\S$\ref{sec:2cygni}); commissioning observations that resolve 2~Cygni for the first time, Keck/NIRC2 follow-up imaging, and direct spectroscopic measurements ($\S$\ref{sec:obs}); the modular data analysis pipeline developed for iLocater's acquisition camera systems ($\S$\ref{sec:analysis}); and results from astrometric, photometric, spectroscopic, and other analyses ($\S$\ref{sec:results}). Concluding remarks are provided in $\S$\ref{sec:conclusions}.

\begin{table}
    \centering
    \begin{tabular}{c|ccc}
    \hline
%    \hline
    Parameter &  Value   & Units & Reference \\
    \hline
%    \hline
      RA     & 19 24 07.58     &  hh:mm:ss  & Gaia EDR3 \\
      DEC    & +29 37 16.81    &  deg:$\farcm$:$\farcs$  & Gaia EDR3 \\
      PM$_{\rm RA}$  & $12.235 \pm 0.101$   &  mas/yr   & Gaia EDR3 \\ 
      PM$_{\rm DEC}$ & $11.229 \pm 0.123$   &  mas/yr  & Gaia EDR3 \\
      $\pi$  & $3.6239 \pm 0.1362$  &  mas   & Gaia EDR3 \\ 
      $U$    & 4.16               & mag &  \cite{reed_03}  \\
      $B$    & $4.851 \pm 0.014$  & mag &  \cite{hog_00}  \\
      $V$    & $4.977 \pm 0.009$  & mag &   Gaia DR3   \\
      $G$    & $4.932 \pm 0.003$  & mag &   \cite{hog_00}   \\
      $J$    & $5.151 \pm 0.017$  & mag &  \cite{cutri_03}  \\
      $H$    & $5.270 \pm 0.033$  & mag &   \cite{cutri_03}  \\
      $K$    & $5.281 \pm 0.021$  & mag &  \cite{cutri_03}   \\
      \hline
%      \hline
    \end{tabular}
    \caption{Observational properties of the 2 Cygni system. Gaia Early Data Release 3 (DR3) is from the \cite{gaia_dr3_21}.}
    \label{tab:2cyg}
\end{table}

\section{2~Cygni BACKGROUND}\label{sec:2cygni}
%We attempt a comprehensive literature search into the nature of 2~Cygni. 

2~Cygni is a naked-eye star system ($V=4.98$) with Table~\ref{tab:2cyg} listing its observational properties. 2~Cygni is identified as a binary in the Catalog of Components of Double and Multiple stars (CCDM) by \citet{dommanget_02}, with original discoverer WRH (R. H. Wilson). However, \citealt{wilson_50} identifies the star as a ``single." Further, the components of 2 Cygni are listed as being separated by 1.58 seconds in right ascension and 16.8 arcseconds in declination. At the distance of 2 Cygni ($d=276\pm11$ pc), this separation corresponds to 8000 au in projection, raising questions about their association. Our observations reveal a companion star at $\theta = 70$ mas that remains to be explained. The neighboring star listed in CCDM at 29$\arcsec$ separation may be aligned by chance and was outside of the field of view of both iLocater and follow-up NIRC2 AO imaging.

A further interest in the 2 Cygni system is its nature as a young ($36.6 \pm 0.5$ Myr) ``run-away" star; \citet{tetzlaff_11} characterize 2 Cygni as a run-away star with a peculiar velocity of $22.9 \pm 2.9$ km/s ($19.4$ km/s tangential to the sky). This speed places it above the peak in the measured three-dimensional space velocity histogram of peculiar stars. Studying the multiplicity of run-away star systems can help to distinguish between event scenarios thought to explain their existence, such as binary supernovae and dynamical ejection from dense young clusters \citep{blaauw_61,poveda_67}.

2~Cygni is also known as HD 182568, HIP 95372, HR 7372, SAO 87159, GC 26785, BD$+$293584, CCDM J19241+2937AB, and 2MASS J19240757+2937169. 2 Cygni has been identified as a probable, long-period astrometric binary by \citealt{turon_93}, \citealt{dommanget_02}, \citealt{wielen_00}, and \citealt{eggleton_08}. However, its true multiplicity is currently unknown.
%and the originally credited discoverer, \citealt{wilson_50}, actually identifies the system as a single.

%Bowler 2016 PASP mentions an imaged brown dwarf around HR 7372, with a citation of Lowrance et al. 2000, but Lowrance et al. 2000 found a brown dwarf around HR 7329.
%\url{https://doi.org/10.1088/1538-3873/128/968/102001}  PASP
%\url{https://doi.org/10.1086/309437} ApJ
% FK6 Catalog: I/264

% Note: Turon+ 1993 is the Hipparcos Input Catalog. Perryman+ 1997 is the Hipparcos Catalog (after the Hipparcos mission). 

\begin{table*}[t]
    \centering
    \begin{tabular}{c|c} %p{0.25\textwidth}}
      \hline
      Reference                 & Comments \\
      \hline
       \citealt{wilson_50}      & single star \\
       \citealt{hoffleit_jaschek_91} & multiple star with reference to \citealt{worley_78} \\
       \citealt{turon_93}       & two components\\
       \citealt{worley_97}      & discoverer codes referenced by \citealt{dommanget_02}\\
       \citealt{hipparcos_97}     &   one component\\
       \citealt{wielen_00}        &   proper motion indicated binary\\
       %Lowrance et al. 2000       &    \\
       \citealt{dommanget_02}      & binary with discoverer ``WRH''\\
       \citealt{makarov_05}       & astrometric binary with discrepant proper motion \\
       \citealt{eggleton_08}      & probable multiplicity of 2; reference to \citealt{makarov_05}\\
       %Bowler 2016                & probable typo  \\
       \citealt{swihart_17}       & assumes 2~Cygni is single \\
       \citealt{gaia_dr3_21}      & RUWE $ = 1.468$            \\
       This Paper                 & 2~Cygni resolved into binary \\
      \hline
    \end{tabular}
    \caption{Select literature references that mention 2~Cygni.}
    \label{tab:literature}
\end{table*}

The Gaia DR3 archive shows a renormalized unit weight error (RUWE) value of 1.468 for 2~Cygni, suggesting that the observed motion and point-spread function (PSF) fitting procedure creates potential problems with a standard astrometric solution for a single star, but the ``non-single-star" flag is set to zero \citep{gaia_dr3_21}. The Hipparcos Input Catalog lists two components, but no astrometric binary flag nor any further notes on the source \citep{turon_93}. Based on available documentation, this information was likely obtained from the CCDM catalog. The Hipparcos Catalog \citep{hipparcos_97} shows a flag of ``system previously identified as multiple in HIC,'' but then lists only one component, as does the update in 2007 \citep{van_leeuwen_07}. The \citealt{wielen_00} FK6 catalog lists 2~Cygni with flags: ``RV may be variable, or composite spectrum or other weak indications of binary or suspected planet'' and ``delta mu binary'' from proper motion differences. 

% Hipparcos: 1993: (I/196)
% Hipparcos 1997: (I/239)

\citealt{eggleton_08} use a number of catalogs to list multiplicities of Hipparcos stars. 2 Cygni (HR 7372) has an ``estimated most probable multiplicity'' of $n=2$ with a flag stating: ``is for an entry that has been identified as a probable astrometric binary by MK, ...''. MK refers to \citealt{makarov_05} who lists ``2Cyg'' in their table of ``A catalog of astrometric binaries with discrepant proper motions in Hipparcos and Tycho-2," but not in their table of ``A catalog of astrometric binaries with accelerating proper motions in Hipparcos." 

% ET reference:  (J/MNRAS/389/869/)
% MK reference:  (J/AJ/129/2420)

%Doesn't list separations but instead a ``Q0 factor.''

% ``... this presumably has a long period, and so if the system is already known to have a short period (from SB9, GCVS, or another source), we assume the system is triple (at least). If this makes it a triple that is not in the MSC, we write $n=3?$ rather than $n=3$. Quite often it is in the MSC, however, because there is convincing additional evidence.'' 

%Swihart reference: (J/AJ/153/16)

In an attempt to identify calibration stars for optical interferometers, \citealt{swihart_17} attempts to weed out known binaries using \citealt{eggleton_08}, but leaves in ``2Cyg.'' \citealt{eggleton_08} provide a cross link to the \citealt{hoffleit_jaschek_91} Bright Star Catalogue, which lists 2 Cygni (HR 7372) as a multiple with a flag that says it was identified by ``Worley (1978) update of the IDS'' and also leaves the ADS value blank. We assume the Worley (1978) update to the IDS refers to \citealt{worley_78} but are unable to locate 2 Cygni in the Worley report.

 %available at \url{https://catalog.hathitrust.org/Record/003948284}. Specifically v.24pt.6. It is a 192 page searchable book, stars are listed by ADS number and IDS number which is related to RA and dec 1900 as well as "discoverer's number, using abbreviations listed in the IDS". There are no stars at the RA and Dec listed as the IDS number from Simbad (although the formatting is different). IDS refers to Index Catalogue of Visual Double Stars, there are many versions it seems from circa 1960s. Simbad refers to a book version from 1961, which I have not been able to find. An article from 1957 describes the type of data in the catalog (\url{https://iopscience.iop.org/article/10.1086/127078/pdf}).

CCDM by \citealt{dommanget_02} lists two components to the 2~Cygni system. Component B has discover listed as ``WRH.''
%Lists an "Original Index (1976.5) or WDS (1994) identifier, based on equinox 1900" of 19202N2925A, no ADS or BDS number and a flag that says there is a note in "Index". Index is very probably the IDS 1961 book version. Read-me file also refers to "INDEX-WDS (C.E.Worley, 1984)" \citep{worley_84}, which is a book of the Washington Double Star catalog (WDS). The identifier 19202N2925A is formatted considerably different than discoverer abbreviations seen in Worley 1978, and there are no results for searching "19202N2925A", "19202", or "2925". Searching "WRH" in Worley 1978 also has no results. 
WRH references ``discoverer codes" in the Washington Double Star catalog \citealt{worley_97} which lists thirteen R.H. Wilson papers plus 1 unpublished manuscript. %Worley 1996 doesn't provide any cross references to check but searching WRH as a discover found no star that matches the RA and Dec (closest was 9Cyg). Same result with the update, Mason et al. 2001.} Worley 1996 lists 13 papers and 1 ``manuscript'' for WRH.  
However, only one of these papers, \citealt{wilson_50}, explicitly lists 2~Cygni, stating that it is a ``single'' instead of listing component separations as with other targets.

Table~\ref{tab:literature} lists relevant references that informed our literature search. We have checked the Keck, Gemini, MMT, ESO, and NAOJ data archives but did not find AO imaging data of 2~Cygni. To our knowledge, 2~Cygni has not been observed with AO using a large telescope and the true multiplicity of the system is currently unknown.

\section{Observations}\label{sec:obs}

\subsection{AO Imaging}\label{sec:AO}
The 2~Cygni system was first observed by the iLocater SX acquisition camera system during instrument commissioning on UT 2019-Jul-09. Follow-up measurements were then obtained with Keck/NIRC2 AO on UT 2022-July-13 to corroborate the iLocater data. More recently, direct spectroscopic measurements were obtained using the ``Lili'' spectrograph at the LBT on UT 2024-May-17. Table~\ref{tab:log} provides an observing log that summarizes the measurements. Images from each instrument are shown in Figures~\ref{fig:LBT_iLocater}, \ref{fig:Keck_H}, and \ref{fig:Keck_Ks}.

\begin{table*}
    \centering
    \begin{tabular}{ccccccccc}
         \hline
         \hline
        Date (UT)     & Instrument &  Filter       &  Mode                & $N_{\rm frames}$  &  $\Delta t_{\rm int}$ [sec]   & $\Delta t_{\rm tot}$ [sec] & Airmass & $\theta$ [$\arcsec$] \\
       \hline
       2019-July-09   & iLocater SX  &   custom         &  Imaging        &  100    &  0.11    &   11      &  1.10      &   0.9       \\
 %      2019-July-10   & iLocater    &   custom     &  Imaging            &         &          &           &   1.7      &   1.1       \\
       2022-July-13   & NIRC2        &  $H_{\rm cont}$  &  Imaging        &  20     &  10.56   &  211.2    &  1.06      &  0.6        \\ 
       2022-July-13   & NIRC2        &  $K_{\rm cont}$  &  Imaging        &  10     &  10.56   &  105.6    &  1.06      &  0.6        \\ 
       2024-May-17    & Lili         &  custom          &  Spectroscopy   &  100    &  0.5     &  50.0     &  1.02      &  0.7-1.0    \\
       2024-May-17    & iLocater SX  &   custom         &  Imaging        &  1000   &  0.11    &   110     &  1.02      &  0.7-1.0    \\
       \hline
       \hline
    \end{tabular}
    \caption{Observing log for AO measurements of 2~Cygni. iLocater's custom imaging channel senses light within a narrow range of wavelengths ($\lambda=0.927-0.960 \; \mu$m). The number of frames ($N_{\rm frames}$), integration time per frame ($\Delta t_{\rm int}$), and total integration time ($\Delta t_{\rm tot}$) are shown for each data set. Seeing measurements ($\theta$) were made using the LBT DIMM system (within 25$^{\circ}$ of target) and Keck MASS/DIMM system.} 
    \label{tab:log}
\end{table*}

%The $H_{\rm cont}$ sequences included 20 exposures consisting of 200 coadded frames with 0.0528 seconds per coadd.
%The $K_{\rm cont}$ sequence included 10 exposures consisting of 200 coadded frames with 0.0528 seconds per coadd.  

\subsubsection{iLocater Imaging}\label{sec:iLocater_ao}
AO imaging observations were performed using iLocater's fine guiding camera (FGC). The FGC uses an Andor Zyla 4.2 Plus high-speed CMOS camera with $6.5 \; \mu$m pixels ($2048 \times 2048$ array). The FGC offers high frame rate imaging to acquire centroid data that helps characterize residual vibrations, tip/tilt correction, and improve fiber coupling performance \citep{abechter_20_smf,crass_21}. Wavelengths in the narrow range $\lambda = 0.927 - 0.960 \; \mu$m are sent to the FGC through the combination of two dichroics, one for the AO system wavefront sensor and another internal to the acquisition camera that diverts light to the spectrograph. 

The SX FGC design offers a field of view of $8 \times 8$ arcseconds with a plate scale of $0.61 \pm 0.08$ arcseconds / mm ($3.97 \pm 0.52$ mas/pix). Images are by default recorded with a changing parallactic angle due to the binocular mounting of each telescope primary mirror. No internal K-mirror systems are used to compensate for field rotation. Instead, the changing parallactic angle is compensated for in software. This feature may be used in the future for high contrast, angular differential imaging measurements to suppress speckles \citep{marois_08}.  A precise atmospheric dispersion corrector compensates for chromatic refraction to improve delivered imaging quality and maximize SMF injection efficiency into the spectrograph. Efforts to precisely calibrate the plate-scale, FGC north-east orientation, and establish rotation angle relations are on-going. Further details of the SX acquisition camera system design and commissioning activities are described in \cite{crass_21}. 

\subsubsection{NIRC2 Imaging}\label{sec:nirc2_ao}
Follow-up measurements using NIRC2 consisted of a sequence of undithered near-infrared exposures in position angle mode using the Keck-2 AO system. A subarray mode of 512$\times$512 pixels was used to offer short integration times. Given the extreme brightness of 2~Cygni, the $K_{\rm cont}$ and $H_{\rm cont}$ filters were used to avoid saturation. The candidate companion star was recovered in both data sets (Fig.~\ref{fig:Keck_H}, Fig.~\ref{fig:Keck_Ks}). Since observations of 2~Cygni occurred during execution of another scientific program, no PSF calibration stars were observed. Images were processed using standard techniques for interpolation of hot pixel values, flat-fielding, and precise image combination \citep{crepp_14_trends}. Photometric and astrometric analysis was performed on the combined images from each filter.

% Several calibration binaries were observed in an attempt to calibrate the FGC; however, prevailing weather prevented the observation of binaries with sufficiently well-characterized orbits.

\subsection{Spectroscopy}\label{sec:spectroscopy_obs}

2~Cygni was observed using a miniature spectrograph named \emph{Lili}, which stands for ``Little iLocater," that we recently installed at the LBT as part of an experiment to characterize the end-to-end performance of the iLocater light path (Harris et al. 2024, submitted). \emph{Lili} covers the iLocater spectral passband ($\lambda = 0.97-1.31 \; \mu$m) with a resolving power of approximately $R \approx 1500$ at a sampling of two pixels per resolution element. As \emph{Lili} is a test spectrograph, only one hardware mode is available: a triple-stacked volume phase holographic grating provides access to two spectral orders that are imaged onto a C-RED 2 camera. The C-RED 2 detector is liquid cooled to -40$^{\circ}$C to minimise dark current. See Harris et al. 2024 for further information on \emph{Lili}. 

The 2~Cygni system was observed during twilight on UT 17-May-2024 at an elevation of approximately 85 degrees; 100 frames were recorded for 2~Cygni~A and 2~Cygni~B each. Seeing was estimated with the LBT DIMM to vary between 0.7 - 1.0 arcseconds. During observations, the detector gain was set to high with an integration time of 0.5 seconds. Data was recorded using the First Light Vision Software program and saved in .FITS format. With the target being resolved by the SX acquisition camera, the instrument SMF was accurately aligned to each stellar component separately using the fiber back-illumination method described in \citet{crass_21}. Significant care was taken to avoid spectral cross-contamination between the two sources: the iLocater SX acquisition camera is capable of centering and maintaining fiber alignment to within $\approx10$ mas, a small fraction of the $\theta=97\pm9$ mas angular separation of the sources during the May 2024 observing run \citep{abechter_20_smf}.

\section{Data Analysis}\label{sec:analysis}

\begin{figure*}
    \centering
    \includegraphics[trim=1.18cm 0.31cm 1.10cm 0.31cm,clip=true,width=2.34in]{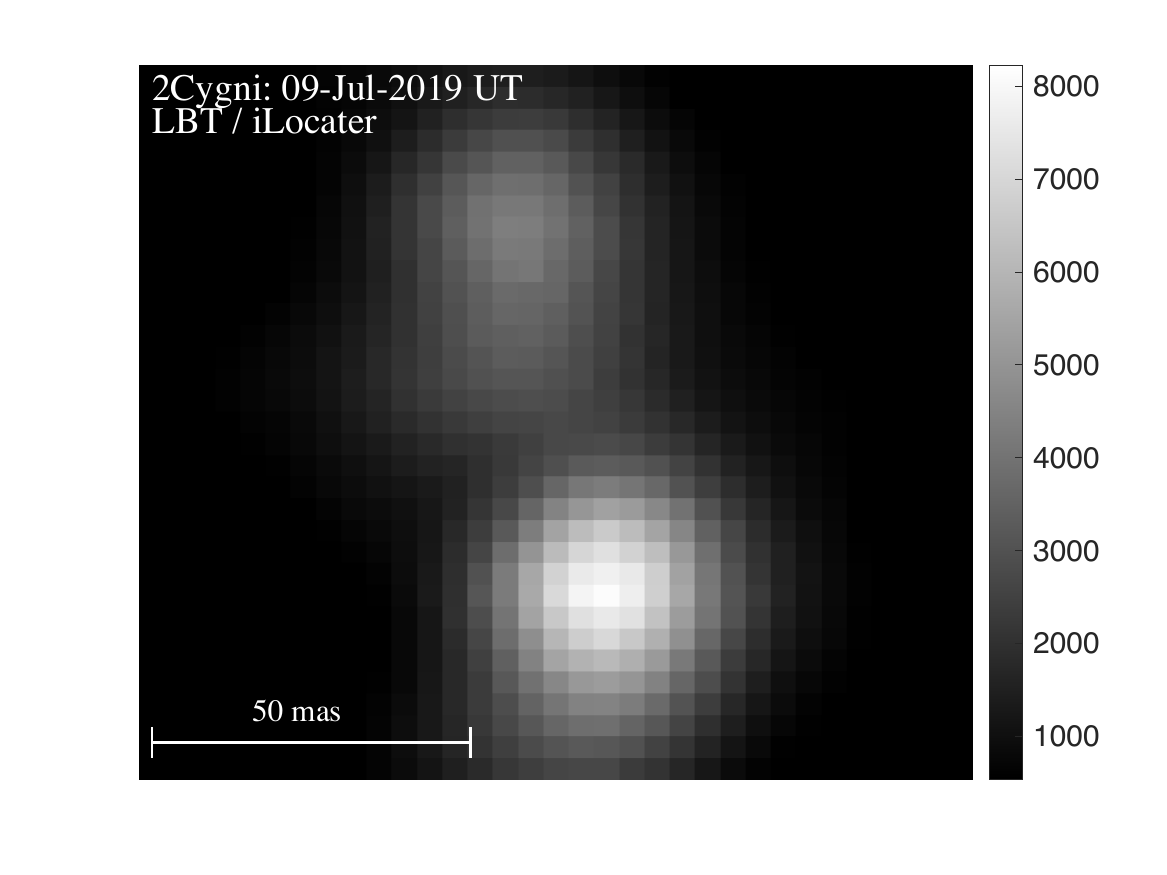}
    \includegraphics[trim=1.18cm 0.31cm 1.10cm 0.31cm,clip=true,width=2.34in]{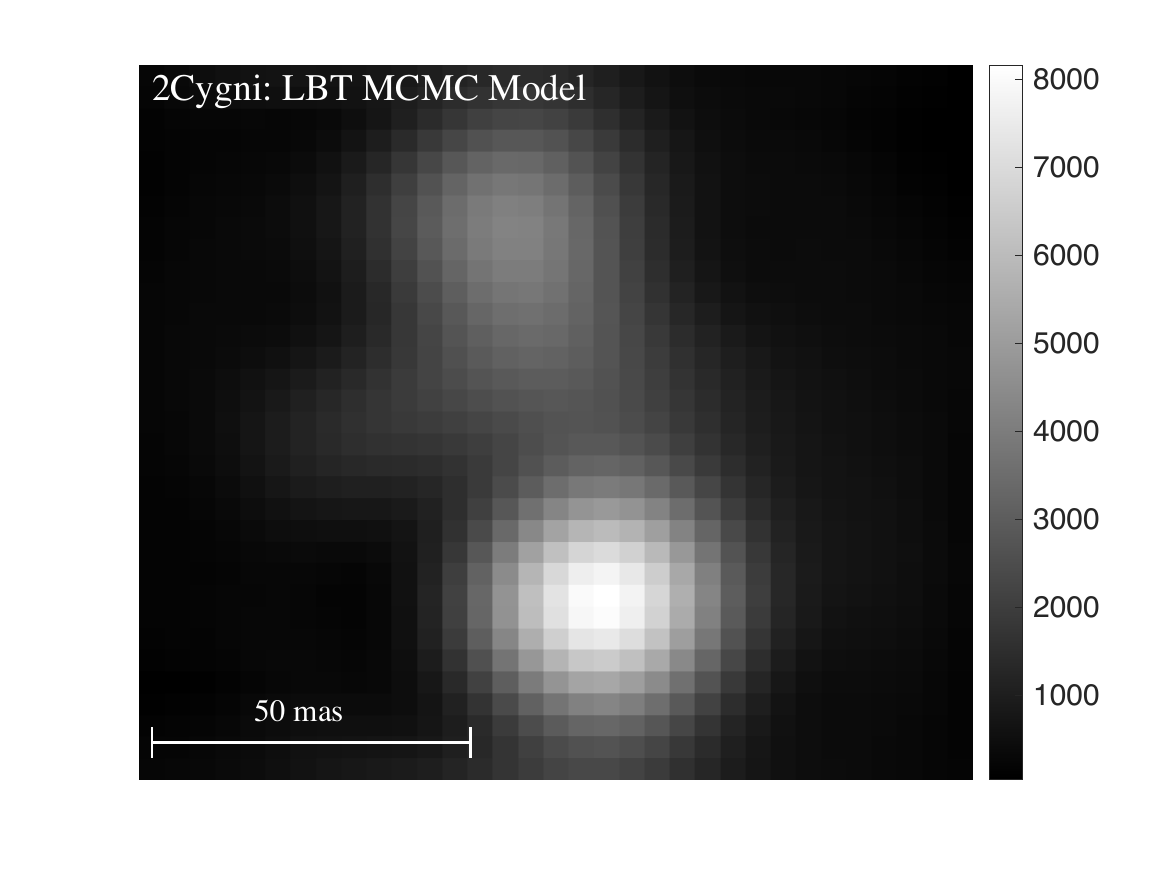}
    \includegraphics[trim=1.18cm 0.31cm 1.10cm 0.31cm,clip=true,width=2.34in]{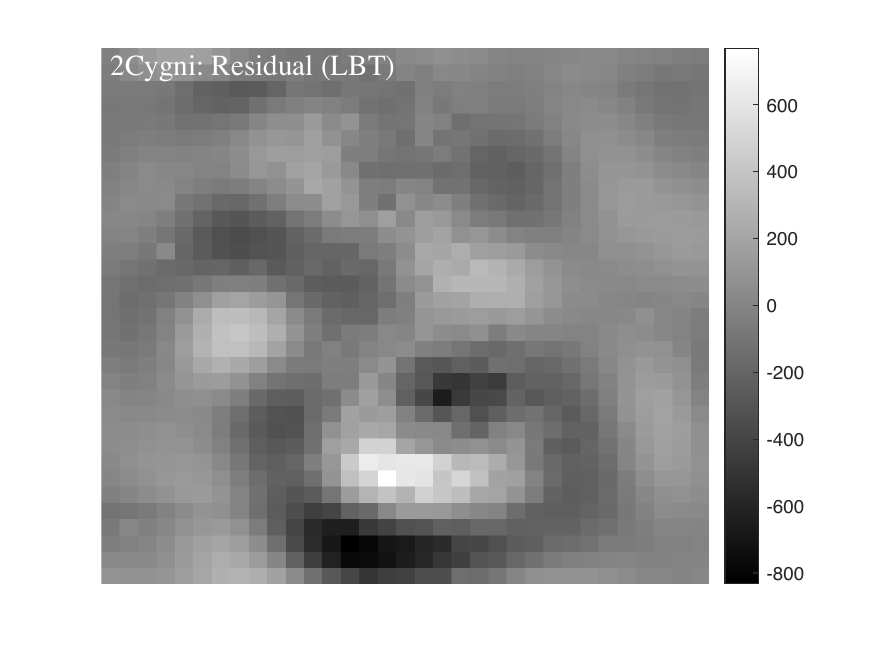}
    \caption{Discovery image recorded with the iLocater SX Acquisition Camera system on 09-July-2019 (left), MCMC model using Zernike polynomials (middle), and intensity residuals (right). North-east orientation has not been calibrated.}
    \label{fig:LBT_iLocater}
\end{figure*}

\begin{figure*}
    \centering
    \includegraphics[trim=1.18cm 0.31cm 1.18cm 0.31cm,clip=true,width=2.34in]{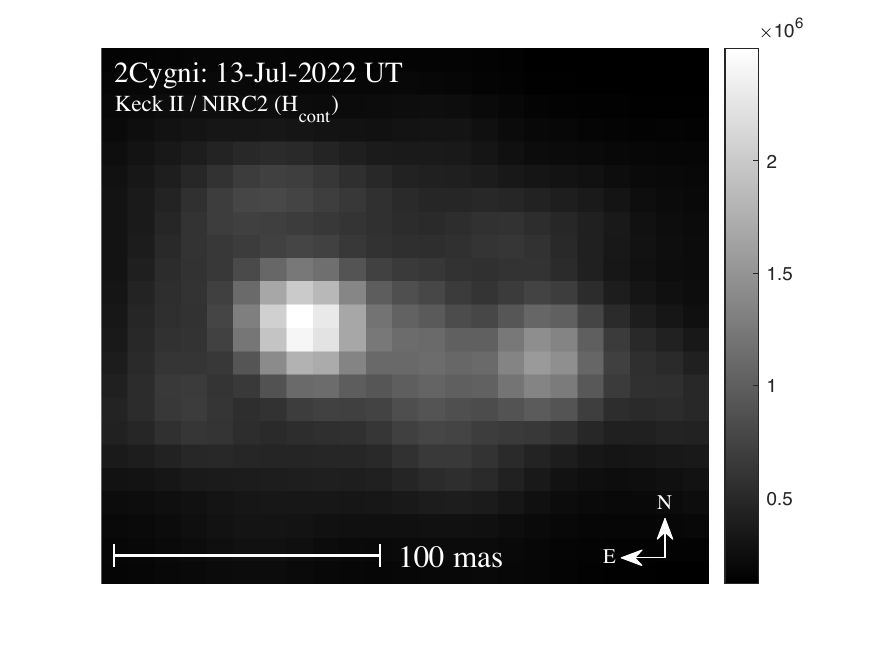}
    \includegraphics[trim=1.18cm 0.31cm 1.18cm 0.31cm,clip=true,width=2.34in]{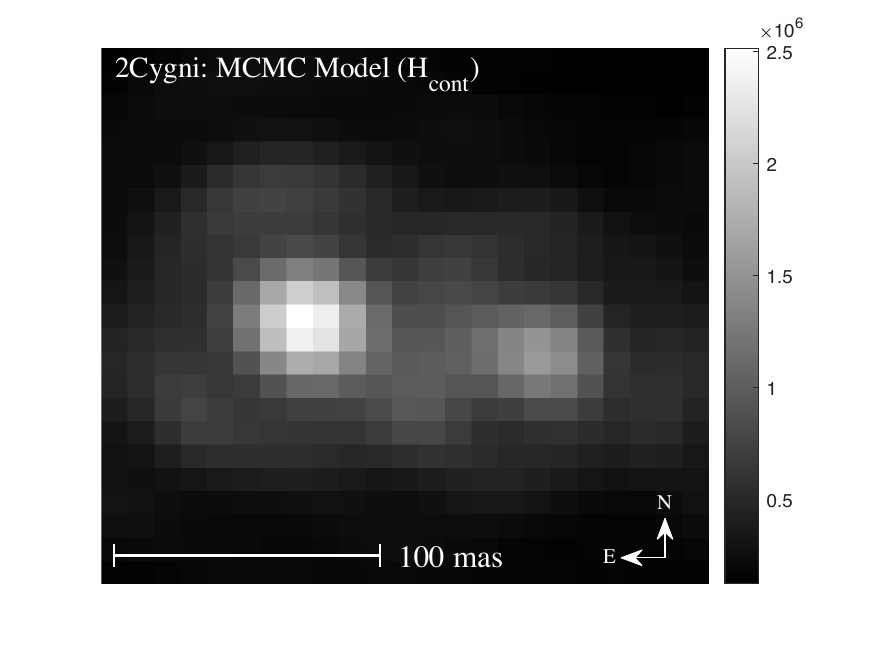}
    \includegraphics[trim=1.18cm 0.31cm 1.18cm 0.31cm,clip=true,width=2.34in]{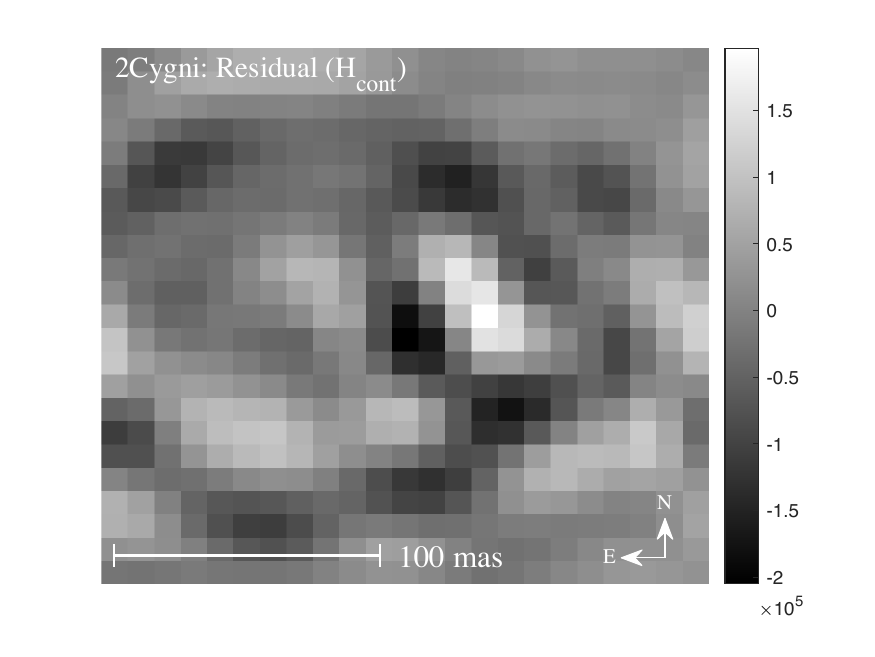}
    \caption{High resolution follow-up images of 2~Cygni taken in the $H_{\rm cont}$ filter using Keck/NIRC2. The three panels show processed data (left), MCMC model (middle), and intensity residuals (right).}
    \label{fig:Keck_H}
\end{figure*}

\begin{figure*}
    \centering
    \includegraphics[trim=1.18cm 0.31cm 1.18cm 0.31cm,clip=true,width=2.34in]{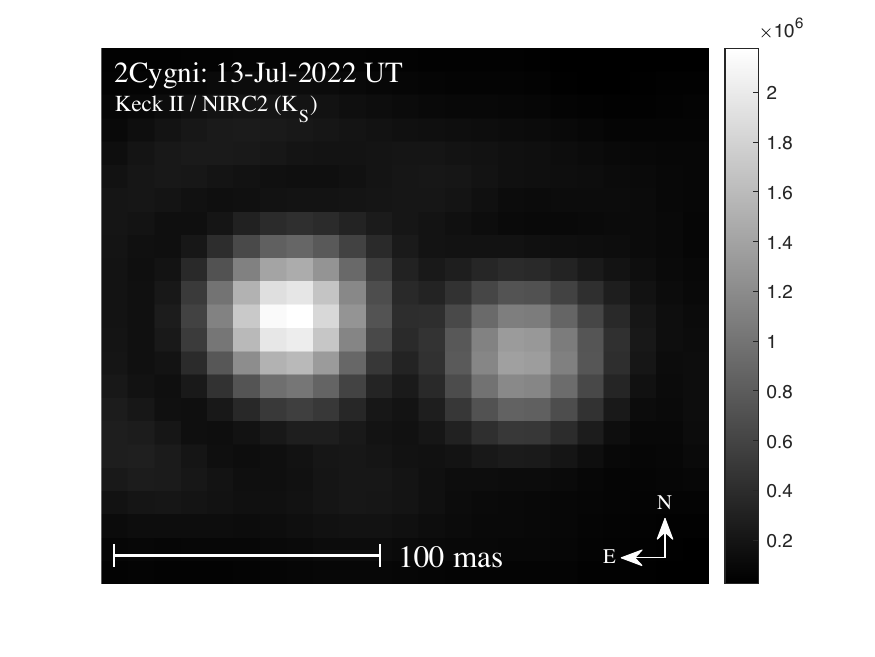}
    \includegraphics[trim=1.18cm 0.31cm 1.18cm 0.31cm,clip=true,width=2.34in]{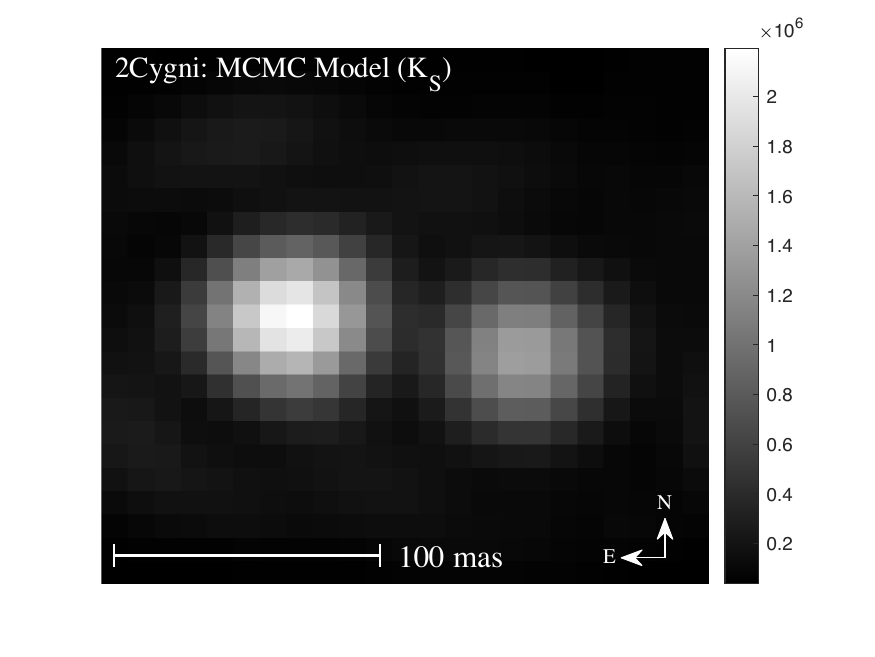}
    \includegraphics[trim=1.18cm 0.31cm 1.18cm 0.31cm,clip=true,width=2.34in]{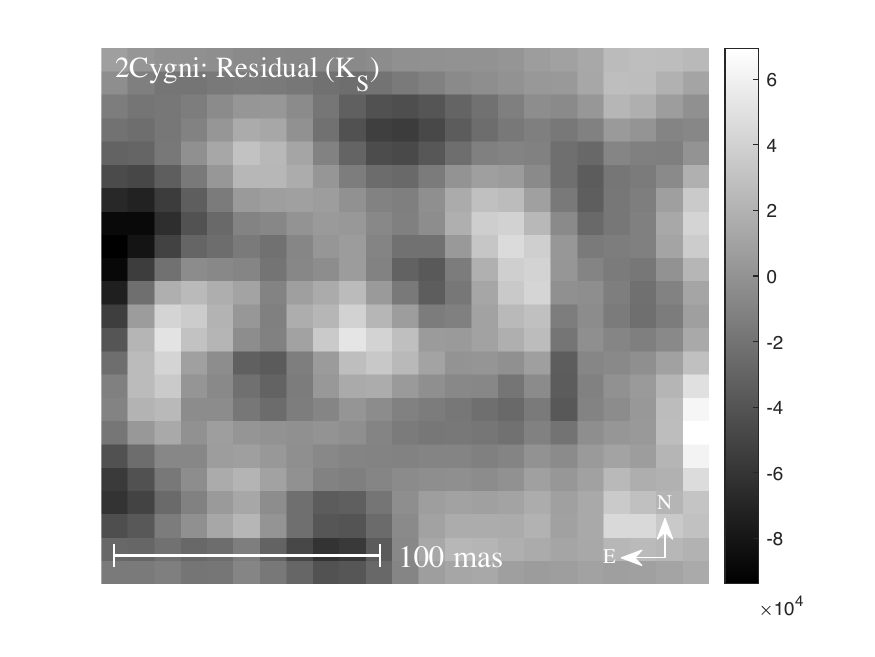}
    \caption{High resolution follow-up images of 2~Cygni taken in the $K_s$ filter using Keck/NIRC2. The three panels show processed data (left), MCMC model (middle), and intensity residuals (right).}
    \label{fig:Keck_Ks}
\end{figure*}

\subsection{Imaging Pipeline}\label{sec:pipeline}
We have developed a semi-automated reduction pipeline that analyzes AO imaging data recorded by iLocater's SX acquisition camera system. The software design is modular so that it can also handle data acquired from the DX acquisition camera system in the future. The pipeline pre-processes (prepares images, files, and directories), cleans (pixel intensity outliers), calibrates (flat-field, background subtract), and derotates, aligns, and combines images recorded through an observing sequence. The pipeline is written in MATLAB and uses methods similar to that of the TRENDS high-contrast imaging program developed for the NIRC2 instrument at Keck \citep{crepp_12,crepp_13,crepp_14_trends}.

Images are first collated and conditioned based on user-defined input values that specify working directories and ancillary information about the data set. Uploaded files are made available for the user to qualitatively assess AO correction and select frames. Standard methods are used for background subtraction, hot pixel cleaning, flat-fielding, etc. 

Telemetry from the telescope control system is used to orient frames based on the evolving parallactic angle throughout the course of an observing sequence. Frames are then co-aligned using a Fourier method that is accurate to the sub-pixel level \citep{guizar_08,crepp_11,pueyo_12}. Finally, the frames are median combined to prepare for analysis. Future versions of the pipeline will incorporate the ability to perform angular differential imaging to search for companions that may be initially hidden by speckle noise. 

\subsection{PSF Fitting}\label{sec:zernikes}

Contamination from the constituent PSF's of closely separated stars can bias the interpretation of photometry and astrometry data. Given that the angular separation of the 2~Cygni binary is comparable to the diffraction-limited spatial resolution of the LBT AO system, we did not attempt aperture photometry. Instead, a Bayesian statistical method was developed and used to account for overlapping PSF's and self-consistently extract relative photometry and astrometry information. The method is similar to \citet{bechter_14}, except that the PSF's are fitted using a physics-based model of aberrations.

We developed an imaging model to fit intensity measurements obtained for each (aligned and median-combined) AO data set. Markov Chain Monte Carlo (MCMC) computations were used to explore the model fit parameter space by numerically calculating a likelihood function for a given set of variables. The Metropolis-Hastings algorithm was implemented to find optimal model parameters and create posterior distributions in order to derive rigorous uncertainties \citep{metropolis_53,hastings_1970}.
 
A Bayesian likelihood function, $\mathscr{L(\phi)}$, is constructed based on model parameters $\phi(x_j,y_j,F_j,S,...)$ that include centroid positions ($x_j,y_j$), peak intensity values ($F_j$), the sky background ($S$), and several nuisance parameters which are described below. The index $j=\{1,2,..\}$ represents each individual star (binary in this case). The likelihood function is evaluated by comparing the intensity of the model at each pixel $i$ to recorded images at the same pixel,
\begin{equation}
    \mathscr{L(\phi)}=\Pi_i \: (2 \pi \sigma_i^2)^{-1/2} \: \exp\left(-\frac{\Delta I_i(\phi)^2}{2 \sigma_i^2}\right),
\end{equation}
where uncertainty in the measured intensity, $\sigma_i$, is assumed to be dominated by photon-noise and given by a Poisson distribution which we approximate as Gaussian. 

Using the Metropolis-Hastings algorithm, we compute the natural logarithm of the likelihood to explore the high-dimensional parameter space of model variables and avoid numerical precision limitations,
\begin{equation}
\ln \mathscr{L(\phi)}=-\frac{1}{2} \sum_i (\ln (2 \pi \sigma_i^2) + \Delta I_i(\phi)^2 / \sigma_i^2).
\end{equation}
The quantity $\Delta I_i(\phi) = I_{model_i}(\phi) - I_{data_i}$ is found by evaluating the intensity contributed by each star in addition to the sky background,
\begin{equation}
    I_{model_i}(\phi) = I_{i,j=1} + I_{i,j=2} + S,
\end{equation}
where $S$ is assumed to be constant over the narrow field of view of iLocater. 

To account for imperfect AO correction, we model the PSF of each star to accommodate aberrations that would otherwise distort and bias the astrometric and photometric solutions. Given the small angular separation of the 2~Cygni components, we assume isoplanatic conditions: light from each star experiences the same turbulence upon arriving to the telescope and therefore the PSFs are essentially identical. We model both PSF's simultaneously using fully reduced images. 

It is assumed that (quasi-)static aberrations dominate the PSF for integrations significantly longer than an AO cycle. Thus, an optical model that includes low-order aberrations was used to fit each stellar PSF. The central obstruction relative diameter of the LBT and Keck were included to model diffraction for each telescope. The first 14 Zernike polynomials (excluding piston and tip/tilt) were superposed in the pupil plane to create a wavefront phase aberration. This number of free parameters strikes a good balance between model accuracy and computational load. Uniform priors were assumed for all Zernike coefficients over an allowable range of $(-1,-1)$. Uncertainties in fitted parameters are given by 68\% confidence intervals derived from posterior distributions.

%Table~\ref{tab:zernikes} provides a description of each orthogonal Zernike term used in the fit (Zernike 1934). 

\begin{center}
\begin{table*}[ht]
    %\centering
    \begin{tabular}{c|ccccc}
    \hline
    \hline
     Star         &     $m_H$           &   $m_{K_s}$          &     $M_H$             & $M_{K_s}$            &  $H-K_s$ \\
    \hline
    \hline
     2~Cygni~A    &  $5.765 \pm 0.069$  &  $5.782 \pm 0.043$   &  $-1.439 \pm 0.108$   &  $-1.422 \pm 0.094$  & $-0.017 \pm 0.082$ \\
     2~Cygni~B    &  $6.360 \pm 0.071$  &  $6.361 \pm 0.044$   &  $-0.844 \pm 0.110$   &  $-0.843 \pm 0.095$  & $-0.001 \pm 0.085$ \\    
     \hline
    \hline
    \end{tabular}
    \caption{Deblended photometry for the 2~Cygni components using NIRC2. $H$-band data is based on $H_{\rm cont}$ measurements and $K_s$-band data is based on $K_{\rm cont}$ measurements.}
    \label{tab:photometry}
\end{table*}
\end{center}

\begin{center}
\begin{table*}[!ht]
   %\centering
   \begin{tabular}{c|cccc}
   \hline
   \hline
   Instrument & Epoch          & $\rho$ [mas]   & PA [$^\circ$]  &  $\Delta m$ \\
        \hline
    iLocater  & 2019-July-09   &  $70 \pm 9$    &    ---             &   $0.66\pm0.10$  \\
    NIRC2     & 2022-July-13   &  $91 \pm 1$    &   $261.1 \pm 0.4$  &   Table~\ref{tab:photometry} \\
    iLocater  & 2024-May-17    &  $97 \pm 9$      &    ---           &   $0.54\pm0.10$  \\
        \hline
        \hline
        % iLocater: 69.9 \pm 9.3
        % 90.8 +/- 0.8  mas NIRC2 weighted average of Kcont and Hcont
   \end{tabular}
   \caption{2~Cygni astrometry showing angular separation ($\rho$), position angle (PA) measured east of north, and magnitude difference in the iLocater filter. Uncertainty in angular separation for iLocater astrometry measurements is dominated by uncertainty in the plate scale; calibration of iLocater's north-east orientation with changing parallactic angle is on-going.}
   \label{tab:astrometry}
\end{table*}
\end{center}

After using a Markov process to create a phase shape, incorporating the telescope central obstruction, the resulting electric field was then propagated to the image plane using a Fourier transform. Taking the square modulus of the image plane field, the resulting intensity pattern was then duplicated to create two identical stellar PSFs. Different centroid positions, equivalent to tip and tilt variables in the pupil plane, were applied using the Fourier shift theorem. The peak flux of each binary component was applied as a multiplicative factor. On-sky images were truncated to a narrow field of view so that the MCMC algorithm could efficiently explore the stellar PSF's. All free parameters were marginalized over when calculating posterior values. 

Following an initial estimate for each free parameter, the MCMC routine was run to find a global minimum for the modeled image. Several million additional iterations were then performed to mix the MCMC chain, explore the parameter space surrounding the global minimum, and create well-sampled posterior distributions. Algorithm convergence was assessed using the Gelman-Rubin criteria by comparing the variance of the chain to the mean value for each parameter \citep{gelman_92}.

\subsection{Spectroscopy with Lili}\label{sec:spectroscopy_analysis}

Data reduction and calibration of \emph{Lili} measurements followed the steps described in Harris et al. 2024, which we briefly summarize. A custom Python program based on the optimal extraction technique of \citet{horne_86} was developed to rotationally interpolate and extract spectra from the two raw spectral orders measured by \emph{Lili}. A Halogen lamp was used to define the spectral profile. A wavelength solution was computed based upon an Ar lamp spectrum. Gaussian fits were used to measure absorption line centroids and a quadratic fit produced the dispersion solution for each order. The resulting wavelength solution achieved an RMS of 0.05 nm or better. The spectra were continuum normalized using a spline fit, excluding regions affected by tellurics or strong stellar lines. A barycentric correction was applied using the Astropy library to Doppler-shift wavelength features based on Earth's motion around the Sun relative to 2~Cygni. 

Observations of \emph{Vega} and \emph{Arcturus} were used to help validate on-sky measurements recorded for 2~Cygni. However, because the observations were conducted during a shared-risk mode (engineering time) with a new instrument, we were unable to observe a telluric reference of similar spectral type in the same part of the sky at the same time as 2~Cygni; thus spectra have not been corrected for telluric features. Nevertheless, we are able to use the spatially resolved spectra to compare each stellar component to one another, as well as to theoretical atmospheric models, in an effort to classify spectral types and constrain effective temperature and surface gravity.

\section{Results}\label{sec:results} 

\begin{figure}[h!t]
    \centering
    \includegraphics[height=2.68in]{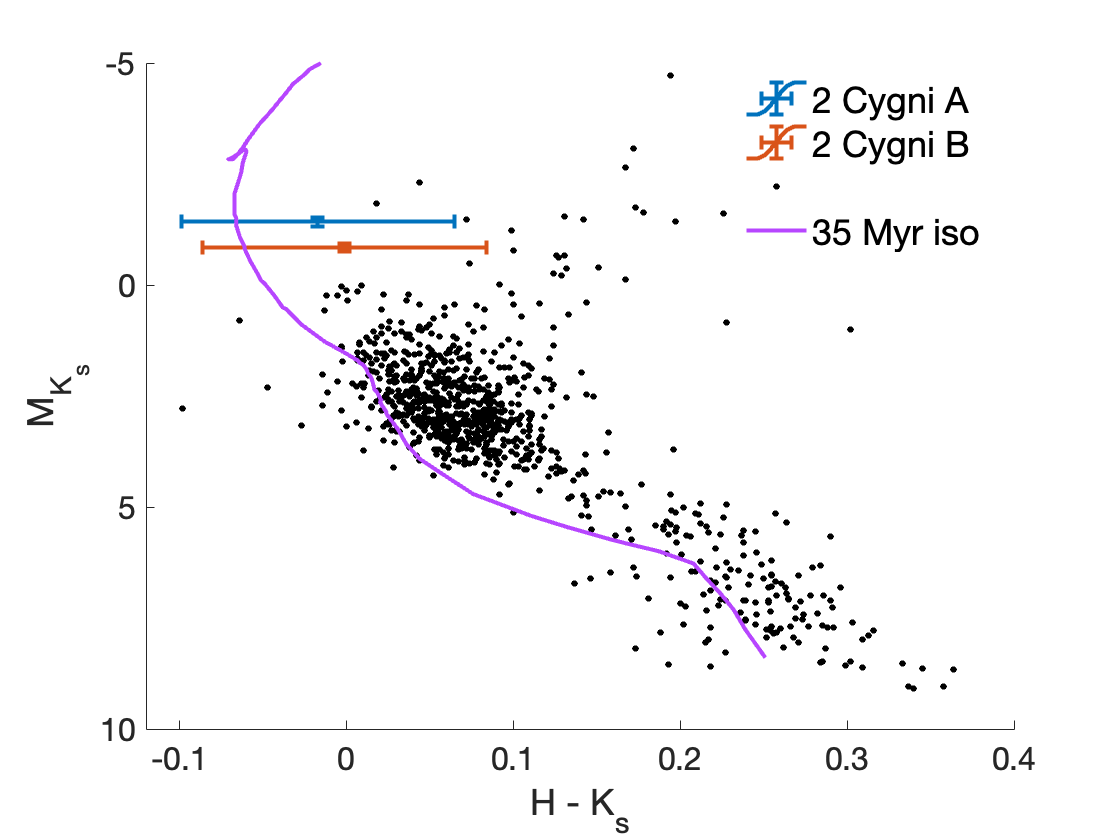}
    \caption{HR-diagram showing the 2~Cygni components. Over-plotted are 1000 stars within 200 pc randomly drawn from the 2MASS catalogue \citep{cutri_03}. A solar-metallicity 35 Myr isochrone from \citet{girardi_02} is shown for comparison.}
    \label{fig:HR_diagram}
\end{figure}

\subsection{Photometry and Astrometry}\label{sec:photometry}

\begin{figure}[h!t]
    \centering
    \includegraphics[height=2.68in]{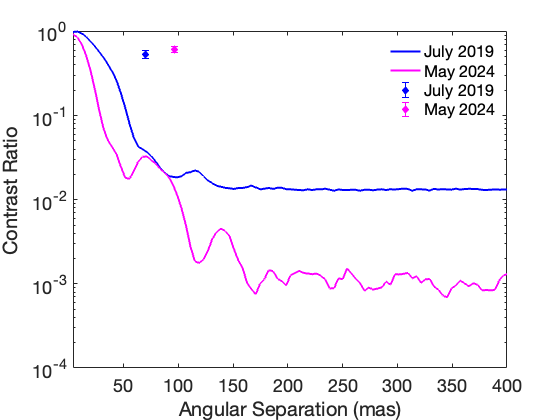}
    \caption{Contrast curves showing the achieved dynamic range for each AO imaging epoch using iLocater. A deeper exposure, lower airmass, and improved instrument calibration offer better contrast for the May 2024 observing run compared with July 2019. The angular separation and flux ratio of 2~Cygni~B is shown for reference.}
    \label{fig:contrast}
\end{figure}

Figures~\ref{fig:LBT_iLocater}, \ref{fig:Keck_H}, and \ref{fig:Keck_Ks} show model images of 2~Cygni compared to data obtained for each observing configuration. Differenced image residuals have RMS values of 2.3\%, 2.4\%, and 1.2\% for the iLocater data (33$\times$33 pixels), Keck $H_{\rm cont}$ and $K_{\rm cont}$ data (22$\times$22 pixels) respectively, indicating a good fit using the MCMC method ($\S$\ref{sec:zernikes}). We use results from the PSF fitting procedure to analyze relative photometry and astrometry of the candidate companion. Figure~\ref{fig:contrast} shows contrast curves for iLocater AO imaging data recorded during the July 2019 and May 2024 epochs. 

Since 2~Cygni does not have a $Y$-band NIR measurement in the literature to our knowledge, and the iLocater camera system uses a custom filter, we deblend the photometry of 2~Cygni's combined light signal using the NIRC2 measurements. Apparent and absolute magnitudes of each source ($A$, $B$) are estimated in the $H$ and $K_s$ bands (Table~\ref{tab:photometry}). Apparent magnitudes are given by
\begin{equation}
    m_A = m_{AB} + 2.5 \log_{10}(1 + f_B/f_A),
\end{equation}
\begin{equation}
    m_B = m_{AB} + 2.5 \log_{10}(1 + f_A/f_B),
\end{equation}
where $m_{AB}$ is the total system apparent magnitude and $f_A$ and $f_B$ are the relative flux of each source. Given the extreme brightness of the 2~Cygni system, we had to use $H_{\rm cont}$ and $K_{\rm cont}$ filters with NIRC2 AO observations to avoid saturation. Thus, the photometry analysis should be considered as an estimate only, interpreted with the caveat that $H_{\rm cont}$ and $K_{\rm cont}$ measurements were used as a proxy in the absence of broadband $H$ and $K_s$ data. 

Assuming that the candidate companion is associated with 2~Cygni (see $\S$\ref{sec:spectroscopy_results} and $\S$\ref{sec:fap}), we calculate absolute magnitudes for the primary and secondary source accounting for the combined light signal (Table~\ref{tab:photometry}). An HR-diagram of the 2~Cygni components is shown in Fig.~\ref{fig:HR_diagram}. 2~Cygni~B appears to be near the main-sequence in close proximity to 2~Cygni~A, with both components being early type stars. 

We note that the multiplicity fraction (binaries plus higher order multiples) of B-type stars exceeds $\approx$80\% and that the distribution is roughly flat with semi-major axis when plotted on a logarithmic scale (e.g. \"{O}pik's law) \citep{offner_23}. Using the absolute magnitudes of each component, we estimate individual masses of $M_A = 6.85 \pm 0.15M_{\odot}$ and $M_B = 5.90 \pm 0.18M_{\odot}$ using solar-metallicity isochrones from \citet{girardi_02} evaluated at an age of 35 Myr. These values are derived as a weighted average using the mass of each component estimated from individual $H$ and $K_s$ band photometry. Such a high mass ratio, $q=M_B/M_A=0.86\pm0.06$, is fairly uncommon for early-type binary stars, which show a peak in the distribution of $q\approx0.3$ \citep{gullikson_16,moe_17}.

\begin{figure}[h]
    \centering
    \includegraphics[height=2.4in,trim={35mm 15mm 35mm 15mm},clip]{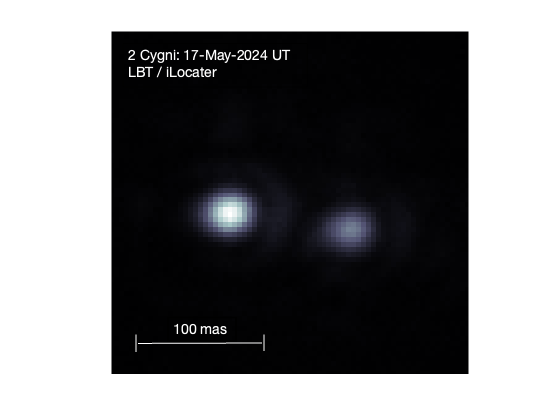}
    \caption{Processed AO image of 2~Cygni from the iLocater SX acquisition camera system recorded in May 2024.}
    \label{fig:AO_May2024}
\end{figure}

\begin{figure*}
    \centering
    \includegraphics[width=0.99\textwidth]{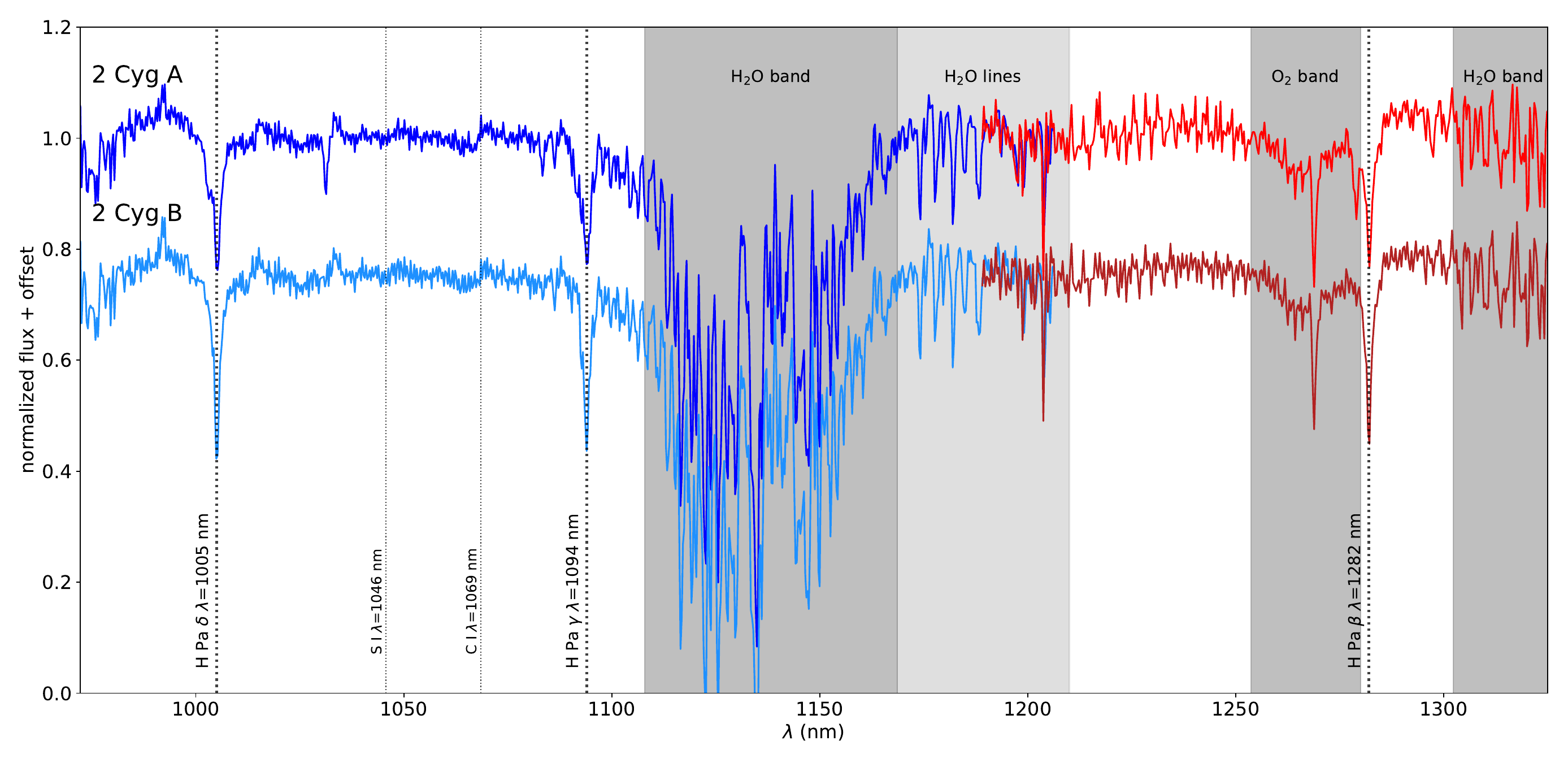}
    \caption{Spectra of the 2~Cygni A and B components recorded with the \emph{Lili} spectrograph. Blue and red colors indicate the two measured orders. The two stars have comparable spectral types and show few absorption lines.}
    \label{fig:lili}
\end{figure*}

\begin{figure*}
    \centering
    \includegraphics[width=0.98\textwidth]{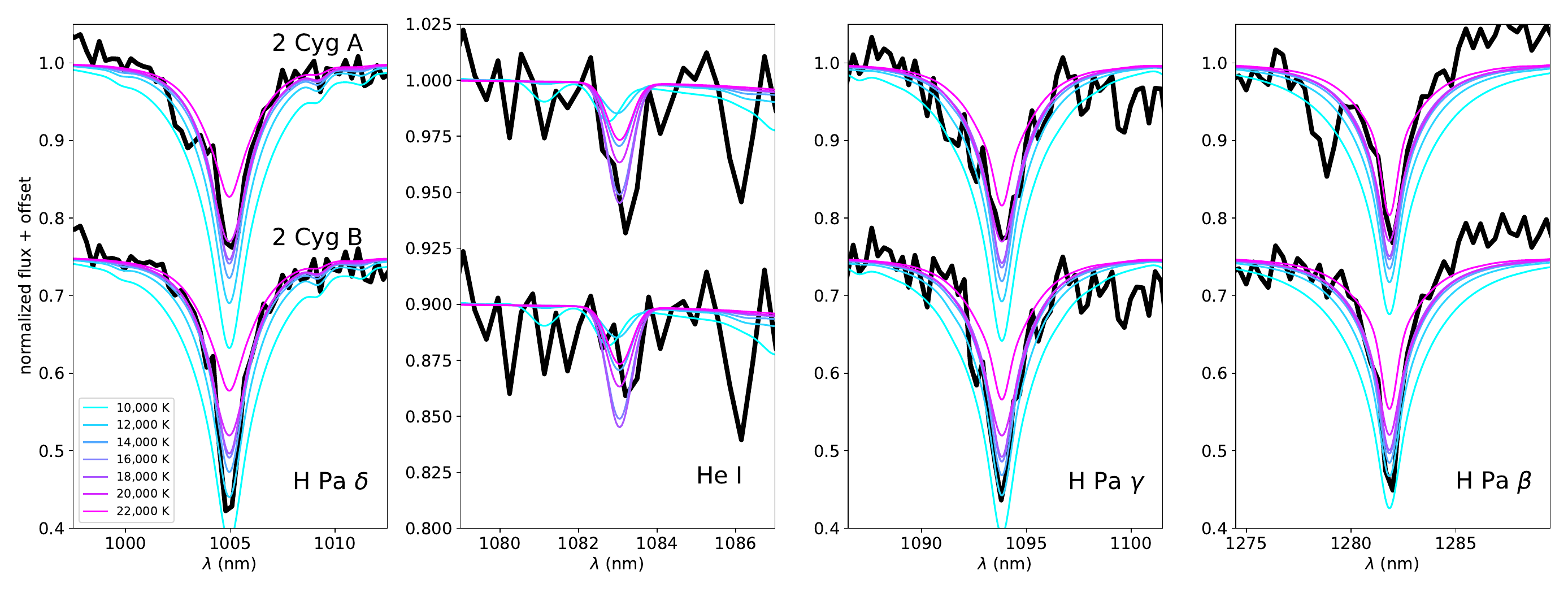}
    \caption{The NIR Paschen Hydrogen line series and neutral Helium compared with BT-Settl models for 2~Cygni~A and 2~Cygni~B. Different effective temperatures are considered in 2000 K increments assuming a surface gravity of $\log(g)=4.0$. In each case, the absorption line depths are consistent with a lower temperature for 2~Cygni~B than 2~Cygni~A.}
    \label{fig:spectrum}
\end{figure*}

\begin{figure}[h]
    \centering
    \includegraphics[height=2.1in]{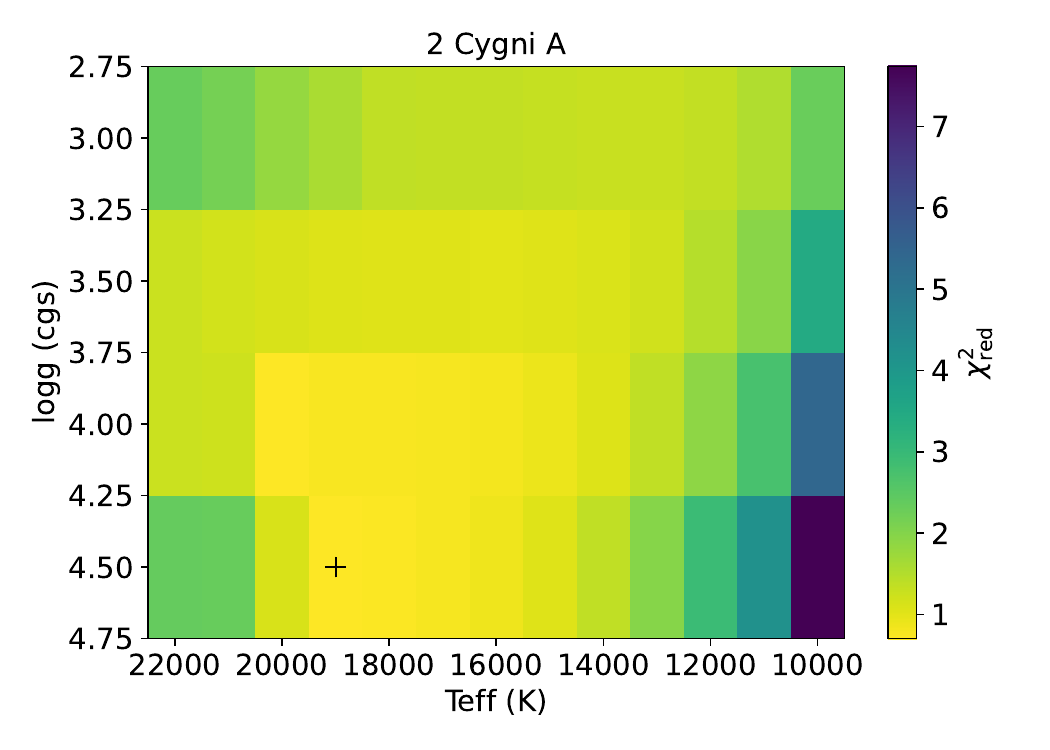}
    \includegraphics[height=2.1in]{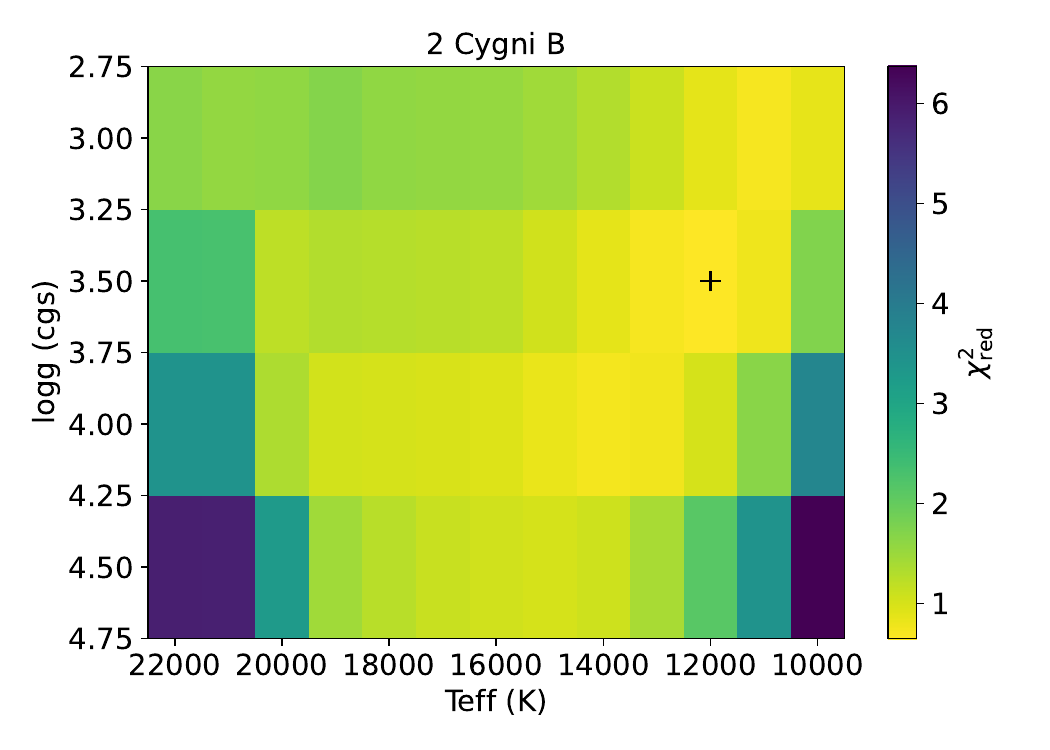}    
    \caption{Grid comparisons for 2~Cygni~A and B when fitting atmospheric models to the full spectrum. A plus sign indicates the location of the lowest RMS value.}
    \label{fig:grid}
\end{figure}

\begin{figure*}[h]
    \centering
    \includegraphics[height=3.9in]{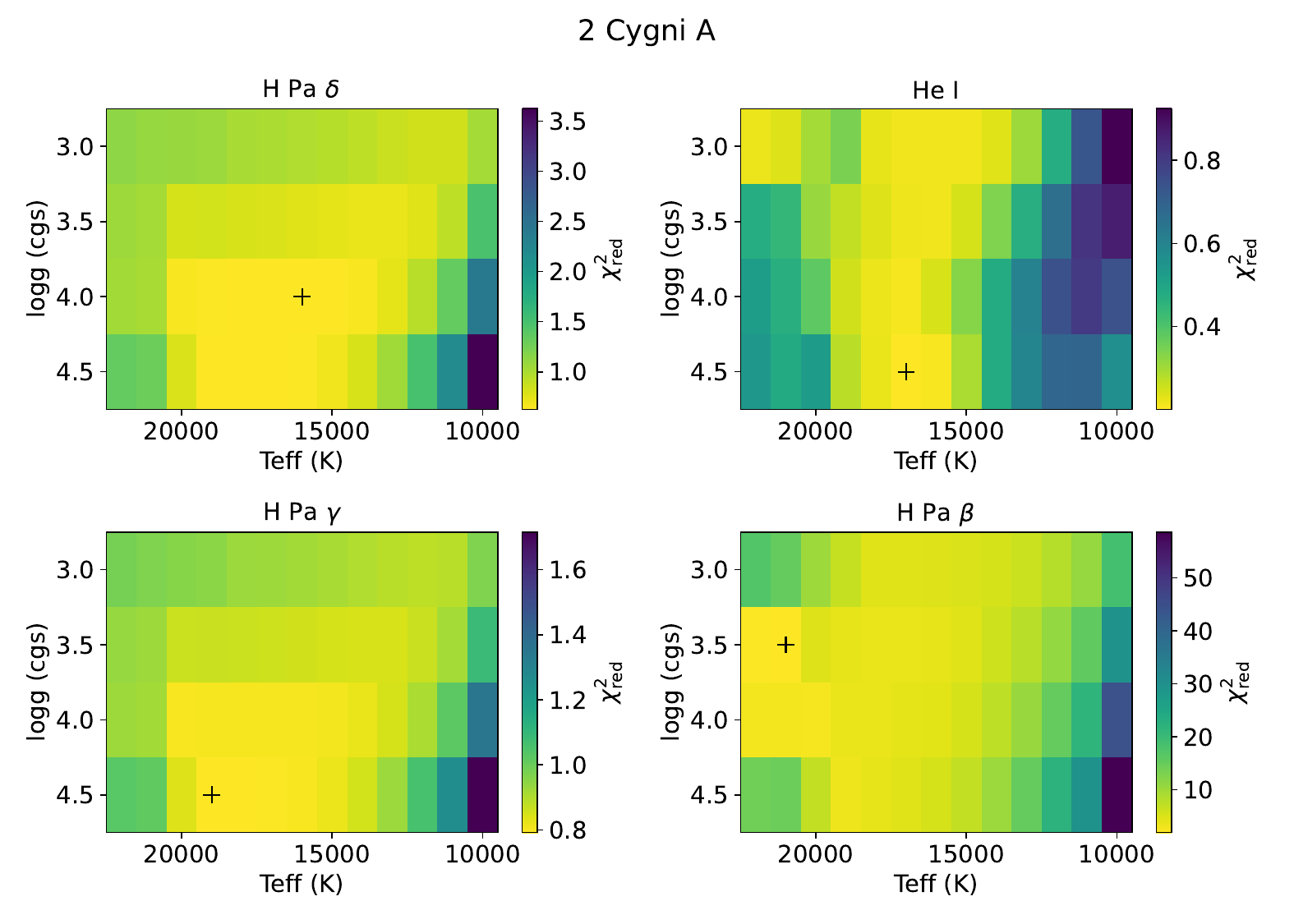}
    \includegraphics[height=3.9in]{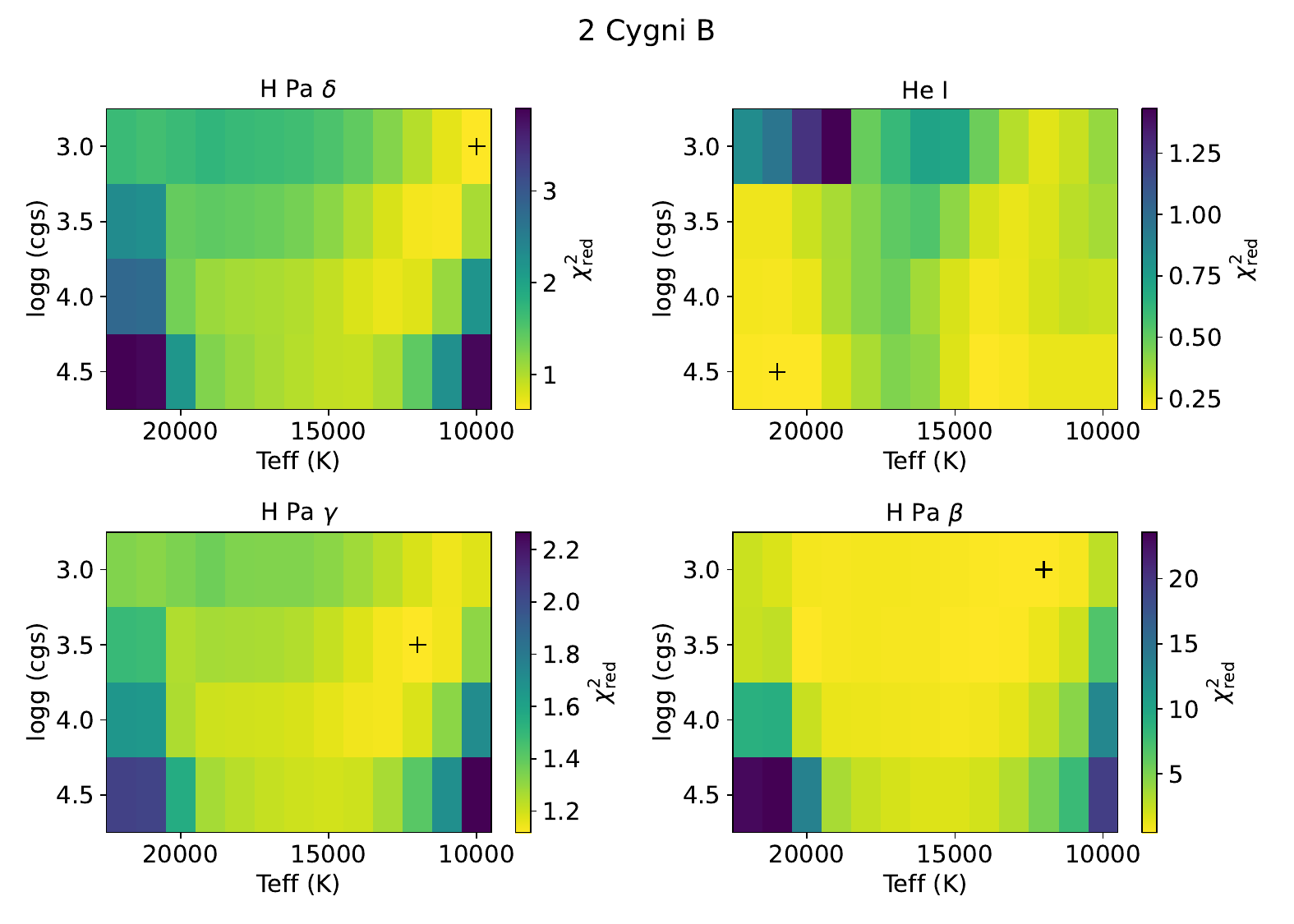}    
    \caption{Grid comparisons for 2~Cygni~A and B when fitting atmospheric models to individual absorption lines. A plus sign indicates the location of the lowest RMS value.}
    \label{fig:linebyline}
\end{figure*}

Astrometric results from each instrument are shown in Table~\ref{tab:astrometry}. Given that iLocater's north-east orientation is not yet calibrated, we estimate the maximum orbital motion that would be possible over a 4.85 year astrometric baseline assuming an edge-on circular orbit. Conservatively using a total system mass of $M_{\rm tot} \approx 13M_{\odot}$ and semi-major axis (equal to the 2019 projected separation, 70 mas $\times$ 276 pc = 19.3 au), the shortest Keplerian orbital period is $P \approx 24$ years. This motion on the sky corresponds to at most an $\approx$89 mas change in angular separation. Given that 2~Cygni's separation has changed by only $27 \pm 13$ mas over this time-frame (Table~\ref{tab:astrometry}), it is plausible that the two stars may be gravitationally associated. Meanwhile, the proper motion of 2~Cygni is $\approx$17 mas/yr corresponding to an 82 mas expected change in separation for an unrelated background source \citep{gaia_dr3_21}. Additional high resolution imaging data with calibrated position angle measurements are required to more definitively assess companionship and attempt to distinguish orbital motion from proper motion and parallactic motion. 

\subsection{Spectral Characterization}\label{sec:spectroscopy_results}
Figure~\ref{fig:lili} shows a comparison of the low resolution spectra extracted for each 2 Cygni component. Blue curves represent the short wavelength order and red curves represent the long wavelength order. Major H$_2$O and O$_2$ telluric bands are indicated. Given the similar structure and features of the spectra, including a dearth of absorption lines, 2 Cygni~B appears to have an early (B-type) spectral classification comparable to 2 Cygni~A.

To estimate the possible spectral contamination experienced by the companion, we generate an AO imaging contrast curve. Contrast was estimated using $1\sigma$ intensity variations as a function of radial separation using the modeled PSF of 2~Cygni~A with the candidate companion PSF removed. As shown in Fig.~\ref{fig:contrast}, the expected contamination of 2~Cygni~B spectra from the nearby 2~Cygni~A primary was $\approx 3$\% ($\theta=97\pm9$ mas) for the May 2024 run. Given the iLocater acquisition camera measured brightness difference of $\Delta m \approx 0.54$ mags, the spatially resolved 2~Cygni~B spectrum can be trusted to $\approx 5$\%. This level of systematic contamination is comparable to spectral variations resulting from not correcting for tellurics.

To further assess the nature of 2~Cygni~A and B, we analyzed the Paschen series of near-infrared Hydrogen absorption lines (H Pa $\delta$, H Pa $\gamma$, and H Pa $\beta$) as well as neutral Helium (He I). Figure~\ref{fig:spectrum} shows each line following continuum normalization for effective temperatures ranging from $T_{\rm eff}=10,000$ K to $T_{\rm eff}=22,000$ K. For comparison, we over-plot BT-Settl spectral models\footnote{\url{http://svo2.cab.inta-csic.es/theory/newov2/}} for different effective temperatures assuming a surface gravity of $\log(g)=4.0$ \citep{allard_11}. For each Hydrogen feature, we find that 2~Cygni~B exhibits a deeper line depth indicating a cooler temperature than 2 Cygni~A. Neutral Helium corroborates this observation by demonstrating consistency through the opposite trend, whereby shallow line depths also point towards a lower temperature for 2~Cygni~B.

As a more comprehensive analysis, we consider a range of surface gravities. Figure~\ref{fig:grid} shows BT-Settl model grid comparisons for the full measured spectrum and Figure~\ref{fig:linebyline} shows a line-by-line analysis \citep{allard_11}. We estimate the uncertainties on the spectral data-points using photon-counting statistics and the known gain and read-noise of the detector, propagated forward through the data reduction process. A reduced chi-squared value is then computed for each model compared to the data, in a narrow window around each absorption line. The best-fitting range of stellar parameters are found from the regions where $\chi^2_{\mathrm{red}}<1$.

At the resolution and sensitivity of the \emph{Lili} measurements, we find that it is possible to discern effective temperatures whereas surface gravity is only loosely constrained. H Pa $\gamma$ and H Pa $\delta$ features are well-fit by theoretical models because they are strong lines without continuum normalization issues in the vicinity. For H Pa $\beta$, the models are less-well fit to absorption lines because the continuum normalization has more difficulty in order 2 (all of the other lines are in order 1) and the blaze efficiency drops rapidly at the end of the order. The He I line is shallow, only barely detected above the noise for 2~Cygni~A and not significantly detected for 2~Cygni~B.

Altogether, this analysis confirms that 2 Cygni~B has a lower $T_{\rm eff}$ than 2 Cygni~A. Meanwhile, model fits for 2~Cygni~A prefer a higher surface gravity than 2~Cygni~B but the results are marginal. Using the full spectrum, 2~Cygni~A is most consistent with $T_A \approx 16,000-20,000$ K and $\log(g) \approx 3.75-4.75$, whereas 2~Cygni~B is most consistent with $T_B \approx 10,000-14,000$ K and $\log(g) \approx 3.00-4.25$. These effective temperatures correspond roughly to spectral types of B3-B4 and B6-B9 respectively \citep{gray_94}, consistent with the combined light visible spectral classification of the 2 Cygni system of B3IV from SIMBAD, given that the signal is dominated at shorter wavelengths by 2 Cygni~A. We can also compare the estimated effective temperatures from spectroscopy to photometry (Fig.~\ref{fig:HR_diagram}). For 2~Cygni~A, we find excellent agreement in that the \citet{girardi_02} models estimate an effective temperature of $T_A = 18,800 \pm 200$ K at an age of $\log(t)=7.55$ ($t=35$ Myr). For 2~Cygni~B, the estimated effective temperature from evolutionary models is lower, $T_B = 17,700 \pm 300$ K, but lies outside of the range indicated by spectroscopic model fits. Thus, further high spatial resolution and high spectral resolution measurements are needed to resolve this tension for the companion temperature.

Finally, we attempted to measure relative RVs of each component using \emph{Lili}. We find that each spectrum shares the same barycentric correction but were unable to detect an offset in Doppler velocity. At the resolution of the \emph{Lili} measurements, we are able to place an upper limit of $\approx \pm 16$ km/s on relative RV between the sources.

The fact that 2~Cygni~B appears to be a star that shares a similar spectral classification as 2~Cygni~A, yet is fainter and lower temperature, argues against the idea of being a chance-alignment foreground or background object. If the 2~Cygni~B candidate were a foreground stellar object, we would expect it to have redder colors than 2~Cygni~A assuming closer distances to the Sun. Instead, 2~Cygni~B has neutral colors and similar spectral type suggesting a similar distance as 2~Cygni~A. If the 2~Cygni~B candidate were a background stellar object, we would expect it to have different colors than 2~Cygni~A, which was shown not to be the case ($\S$\ref{sec:photometry}).

\subsection{False-Positive Analysis}\label{sec:fap}
We use the TRILEGAL galactic model tool from \citealt{girardi_05} to estimate the number density of stars in the direction of 2~Cygni. The model includes thin disk, thick disk, and halo components. The false-positive probability of coincident line-of-site sources is estimated by considering an area of sky with radius $r=90$ mas. Located 6 degrees from the galactic plane, TRILEGAL expects to find approximately 11 sources having $H<7$ within an area of 0.5 square degs centered around 2~Cygni. Assuming a uniform rate of occurrence in this region of interest, the expected number of sources located within 90 mas of 2~Cygni is $\approx 4.3\times10^{-8}$. Given this finding, along with the fact that the secondary has a similar color as 2~Cygni~A, we conclude that the likelihood of the imaged source being an unrelated foreground or background object is low. 

\section{Summary and Concluding Remarks}\label{sec:conclusions}

The first major sub-system of the iLocater instrument, the SX Acquisition Camera, has been installed and commissioned at the LBT. The unit injects starlight directly into single mode fibers for spectroscopy in the wavelength range $\lambda = 0.97 - 1.31 \; \mu$m, while shorter wavelengths ($\lambda= 0.927 - 0.960 \; \mu$m) are diverted to a high-speed camera to record diffraction-limited images for both science and real-time diagnostics \citep{abechter_20_smf,crass_21}. In this paper, we demonstrate the capabilities of this newly-commissioned AO instrument by reporting the serendipitous imaging discovery of a previously unknown binary star system with a separation of only $\theta = 70$ mas. 

We further demonstrate the first end-to-end spectroscopic measurements using the precursor \emph{Lili} spectrograph, which helped to characterize each binary star component (Harris et al. 2024). Broadband measurements at $R\approx1500$ show a similar spectral energy distribution for each resolved star. The estimated range of effective temperatures from Hydrogen absorption features suggests that 2~Cygni is comprised of an early-type B-star and late-type B-star. 

Evidence suggesting 2~Cygni's binary nature consists of multi-epoch, multi-instrument, multi-filter direct imaging detection of a candidate source with relative brightness, colors, and spectrum consistent with companionship. The off-axis source can likely only be detected using AO (or possibly speckle imaging) on a large telescope. The false-positive probability of a nearby foreground or background stellar source located in such close proximity to 2~Cygni with these attributes is less than one in $1/4.3\times10^{-8}=23\times10^6$. 

Representing the second-brightest star in the Cygnus constellation, the naked-eye ($V=4.98$), B-star 2~Cygni appears to be a massive visual binary system. Additional follow-up measurements are needed however to break degeneracies between proper motion and possible orbital motion, determining with certainty whether the candidate is a bona-fide companion. If gravitationally bound, the full orbit of 2~Cygni could be reconstructed in the future with long time-baseline Doppler RV measurements combined with AO imaging. Distinguishing between orbital motion and galactic space motion will be important for precisely tracing the trajectory of 2~Cygni backwards in time. The detection of 2~Cygni~B may impact the parallax measured by Gaia and estimated absolute magnitudes; thus system parameters will need to be updated with any refinements of the distance to 2~Cygni. Finally, it remains to be seen whether a third, more distant, source may be associated with 2~Cygni at tens of arcseconds separation as suggested by \citet{dommanget_02}. 

Nearby companions can impact the evolution of massive stars in binary systems, but only those with orbital periods less than $\approx 1500$ days might exchange mass \citep{sana_12}. The 2~Cygni components are sufficiently separated that mass exchange is unlikely to occur. Many young, massive stars are known to host a close binary companion, but the fraction depends on environment (field, cluster, run-away stars, walk-away stars, etc.) \citep{chini_12}. One way to produce anomalous spatial velocities is through disruption of binary systems via mass transfer, resulting in a lower number of companions; alternatively, ejection from a star cluster, e.g. through interaction with a blackhole, can give rise to peculiar velocities \citep{renzo_19}. Thus, studying the multiplicity and kinematics of early-type systems like 2~Cygni can help to inform galactic formation scenarios by providing context for their evolution. 

% If the explosion of a third star accelerated the 2~Cygni pair to higher than normal velocity, one could speculate that only a tight binary would remain bound.

\section{Acknowledgements}\label{sec:acknowledge}
We thank the many LBTO staff and engineers for their tireless efforts in facilitating the transport, rebuild, integration, and testing of the iLocater SX acquisition camera system. This research is based upon instrumentation work supported by the National Science Foundation under Grant No. 2108603. JRC acknowledges support from the NASA Early Career Fellowship (NNX13AB03G) and NSF CAREER programs (1654125). This research has made use of the Washington Double Star Catalog maintained at the U.S. Naval Observatory. M.C.J. acknowledges support from the Thomas Jefferson Endowment for Space Exploration at Ohio State. This research was supported in part through a service agreement with the Notre Dame Engineering and Design Core Facility (EDCF). We thank David Futa, Matthew Sanford, and Gary Edwards from the Notre Dame physics machine shop for building custom mechanical components. We are grateful for feedback from the iLocater acquisition camera preliminary design review (PDR) and final design review (FDR) committee panel members, including Nemanja Jovanovic, Bertrand Mennesson, Chad Bender, Julian Christou, Olivier Lai, Reed Riddle, Tom McMahon, Douglas Summers, Christian Veillet, and others who offered helpful suggestions and insights that improved the instrument design. Finally, we are very grateful for the support and vision of the Potenziani family and Mr. Ted Wolfe. The LBT is an international collaboration among institutions in the United States, Italy, and Germany. LBT Corporation partners are: The University of Arizona on behalf of the Arizona university system; Istituto Nazionale di Astrofisica, Italy; LBT Beteiligungsgesellschaft, Germany, representing the Max-Planck Society, the Astrophysical Institute Potsdam, and Heidelberg University; The Ohio State University, and The Research Corporation, on behalf of The University of Notre Dame, University of Minnesota, and University of Virginia. Observations have benefited from the use of ALTA Center (\url{alta.arcetri.inaf.it}) forecasts performed with the Astro-Meso-Nh model. Initialization data of the ALTA automatic forecast system come from the General Circulation Model (HRES) of the European Centre for Medium Range Weather Forecasts.

\end{CJK*}
\end{document}